\documentclass{ws-procs9x6}

\newcommand{\Tr}{{\rm Tr}}
\newcommand{\C}{{_{\mathcal C}}}
\newcommand{\beq}{\begin{equation}}
\newcommand{\eeq}{\end{equation}}
\newcommand{\bea}{\begin{eqnarray}}
\newcommand{\eea}{\end{eqnarray}}
\newcommand{\beqn}{\begin{equation*}}
\newcommand{\eeqn}{\end{equation*}}
\newcommand{\bean}{\begin{eqnarray*}}
\newcommand{\eean}{\end{eqnarray*}}
\newcommand{\fslash}{\!\!\!\!/\,}

\begin{document}

\title{\vspace*{-0.5cm}
Progress in nonequilibrium \\quantum field theory
\footnote{\uppercase{P}repared for \uppercase{SEWM}2002. \uppercase{B}ased on an
invited talk by \uppercase{J}.\uppercase{B}.\ and a poster 
by \uppercase{J}.\uppercase{S}.}}

\author{\vspace*{-0.2cm}J\"urgen Berges\footnote{email: j.berges@thphys.uni-heidelberg.de}$\,\,$
and$\,$ Julien Serreau\footnote{email: serreau@thphys.uni-heidelberg.de}}

\address{Institut f\"ur Theoretische Physik der Universit\"at Heidelberg\\
Philosophenweg 16, 69120 Heidelberg, Germany}


\maketitle

\abstracts{\vspace*{-0.2cm}We review recent developments for the description 
of far-from-equilibrium dynamics of quantum fields and subsequent 
thermalization.}
\vspace*{-0.8cm}
\section{Far-from-equilibrium quantum fields}

The advent of relativistic heavy-ion collision experiments as well as 
rapid progress in cosmological observations triggered an 
enormous increase of interest in the dynamics of quantum fields out 
of equilibrium. Much progress has been achieved for systems close to 
thermal equilibrium using \mbox{(non-)linear} response techniques or with 
effective descriptions based on a separation of scales in the
weak coupling limit\cite{Bodeker:2001pa,YaffeMooreProkopec}.
These methods provide an efficient description of the dynamics on 
sufficiently long time scales in their range of applicability.    
Major open questions for our theoretical understanding are currently
highlighted by experimental indications of early thermalization in collision 
experiments starting from extreme nonequilibrium 
situations\cite{Braun-Munzinger:2001mh,Serreau:2001xq}. They
provide an important motivation to study the far-from-equilibrium 
dynamics of quantum fields and subsequent 
thermalization from first principles. 

There are few necessary ingredients to describe far-from-equilibrium 
quantum fields. Firstly, in contrast to close-to-equilibrium field theory,
the initial density matrix can deviate substantially from thermal 
equilibrium. Fully equivalent to the specification of a
density matrix is a description of the nonequilibrium 
initial conditions in terms of correlation functions. 
Secondly, nonequilibrium time evolution involves no other
dynamics than the one dictated by the underlying quantum field theory.
In particular, the dynamics can be conveniently obtained from
the effective action $\Gamma$ for given 
initial correlation functions. Since $\Gamma$ is a Legendre transform 
of the generating functional for connected Green's functions, there 
are very efficient approximation schemes available. 

Practicable and systematic approximations for nonequilibrium
situations may be based on the two-particle irreducible (2PI) 
effective action\cite{Cornwall:1974vz,KadanoffBaym,Chou:1985es}. 
It has been recently demonstrated for 
scalar\cite{Berges:2000ur,Berges:2001fi,Cooper:2002qd}
and fermionic\cite{Berges:2002wr} quantum field theories 
that a systematic coupling--expansion\cite{Cornwall:1974vz} 
or $1/N$--expansion\cite{Cornwall:1974vz,Berges:2001fi,Aarts:2002dj} 
of the 2PI effective action can describe far-from-equilibrium
dynamics and subsequent thermalization without
further assumptions. At lowest nontrivial order these approximations 
contain scattering contributions including off-shell and memory effects.
In particular, the 2PI $1/N$--expansion has been shown to solve the 
problem of an analytic description of the dynamics at nonperturbatively 
large densities\cite{Berges:2002cz,Berges:2002wt}.

A related approach that has been successfully 
applied\cite{Cooper:2002qd} to nonequilibrium quantum field theory
is motivated by truncating Schwinger-Dyson 
equations\cite{Mihaila:2000sr}. It is equivalent 
to an approximation including the complete next-to-leading order 
(NLO) contribution and part of the NNLO graphs 
of the 2PI $1/N$-expansion\cite{Aarts:2002dj}.
Similar to the 2PI effective action technique, promising
approximation schemes may also be obtained in a systematic way
by considering so-called two-particle ``point-irreducible'' (2PPI) 
graphs. In the context of nonequilibrium this has been 
investigated in Ref.~\refcite{Baacke:2002ee}.

These methods provide a systematic and practical tool to go beyond
standard mean-field type approximations\cite{Boyanovsky:vi}, such 
as \mbox{Hartree(--Fock)} or leading order large--$N$.
Although the latter present a useful guide for various problems, they e.g.\
fail to describe exponential damping of correlations
at early times or the late-time dynamics leading to 
thermalization\cite{Berges:2000ur}.\footnote{For somewhat 
improved results using inhomogeneous mean fields see Ref.~\refcite{LOinh}.}
These problems are related to the appearance of an infinite number of conserved
quantities, which are not present in the fully interacting or finite-$N$
field theory. The spurious constants of motion restrict the nonequilibrium
evolution at all times and prevent an approach to thermal 
equilibrium. Although a straightforward task in 
principle, going beyond mean-field type approximations has long been 
a major difficulty in practice: Similar to perturbation
theory, standard approximations based on $1/N$--expansions of 
the one-particle irreducible (1PI) effective action 
can be secular in time. Even for small 
couplings these approximations break down after a short time and do not 
describe thermalization\cite{Bettencourt:1997nf}.  In the following 
we review two-particle irreducible expansion schemes and recent 
nonequilibrium applications.

\vspace*{-0.1cm}
\section{2PI effective action for scalar and fermionic fields}

\begin{figure}[b]
\vspace*{-0.1cm}
 \centering
 \begin{minipage}[c]{.45\textwidth}
  \centering
  \centerline{\epsfxsize=3.5cm\epsffile{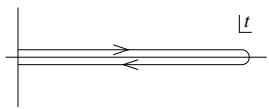}}
 \end{minipage}
 \begin{minipage}[c]{.45\textwidth}
  \centering
  \caption{\label{fig:CTP} Traces involve a {\em finite} 
  closed time path $\mathcal{C}$ suitable for the initial
  value problem with a causal time evolution.}
 \end{minipage}
\vspace*{-0.1cm}
\end{figure}

Consider a theory describing $N$ scalar fields $\varphi_a$
and $N_f$ fermion flavours $\psi_i$ 
with classical action ($a,b=1\ldots N$; $i,j=1\ldots N_f$)
\beq
 S[\varphi;\bar{\psi},\psi] = \int {\rm d}^4 x \Big\{
 \frac{1}{2} \partial_\mu \varphi_a \, \partial^\mu \varphi_a
 +\bar{\psi}_i(x) \,i \partial\,\fslash \psi_i(x)
 + V(\varphi;\bar{\psi},\psi) \Big\} \, .
\label{classaction}
\eeq
Here $V(\varphi;\bar{\psi},\psi)$ parametrizes the interaction part to be
specified later, and \mbox{$\partial\,\fslash \equiv \gamma_\mu \partial^\mu$}
with Dirac matrices $\gamma_\mu$ ($\mu = 0\ldots3$).
All Green's functions of this theory can be obtained from the 
2PI effective action\cite{Cornwall:1974vz}, which is 
parametrized by the connected one- and two-point functions: 
\beqn
 \phi_a (x) \equiv \langle \, \varphi_a (x) \, \rangle \,\,\, , \,\,\, 
 G_{ab} (x,y) \equiv \langle \, {\rm T}_\C \varphi_a (x) \, \varphi_b (y)
 \, \rangle - \phi_a (x)\,\phi_b (y) \, ,
\eeqn
and similarly for the fermionic fields. Here, the field operators are 
time-ordered along a closed time path $\mathcal C$, as depicted
in Fig.\ \ref{fig:CTP}.
For a vanishing fermionic ``background'' 
field the 2PI effective action can be written as\cite{Cornwall:1974vz}
\bea
 \Gamma[\phi,G;D] &=& S[\phi;0,0] + \frac{i}{2} \Tr_\C\ln G^{-1} 
 + \frac{i}{2} \Tr_\C \,G_0^{-1}G \nonumber\\
 &&-i \Tr_\C\ln D^{-1} -i \Tr_\C\, D_0^{-1} D
 + \Gamma_2[\phi,G;D] \, ,
\label{2PIaction}
\eea
where the classical inverse propagator for the scalars is given by
\beq
 i G^{-1}_{0,ab}(x,y;\phi) =
 \left.\frac{\delta^2 S[\phi;\bar{\psi},\psi]}{\delta\phi_b(y)\delta\phi_a(x)}
 \right|_{\bar{\psi}=\psi=0} \\
\eeq
and equivalently for the classical inverse fermion propagator $i D^{-1}_0$.  
The term $\Gamma_2[\phi,G;D]$ in Eq.\ (\ref{2PIaction}) contains all 
contributions beyond one-loop order and can be represented as 
a sum over closed 2PI graphs only\cite{Cornwall:1974vz}, i.e.~diagrams
which cannot be disconnected by opening two propagator
lines. The equations of motion are obtained by extremizing the 
effective action:
\beq
 \frac{\delta \Gamma[\phi,G;D]}{\delta \phi_a(x)} = 0 \,\, ,\,\,\,
 \frac{\delta \Gamma[\phi,G;D]}{\delta G_{ab}(x,y)} = 0 \,\, ,\,\,\,
 \frac{\delta \Gamma[\phi,G;D]}{\delta D_{ij}(x,y)} = 0 \, .
\label{EOM}
\eeq
The resulting equations of motion can be written as differential 
equations suitable for initial value problems 
(for details see~e.g.~Ref.~\refcite{Berges:2001fi}). 
All higher $n$-point functions can then be obtained 
from the generating functional $\Gamma[\phi,G;D]$ at any
time during the evolution. Without further truncation
the effective action contains the complete information.

\section{Systematic approximation schemes}
\label{sect:app}

A major advantage of the 2PI effective action is that it allows
one to obtain suitable approximation schemes for nonequilibrium
problems in a systematic way.    
As an example, we consider first a scalar field theory with 
$O(N)$--symmetric classical action given by Eq.\ (\ref{classaction}) 
with
\beq
\label{potential:largeN}
 V(\varphi)  = \frac{m^2_0}{2} \, \varphi_a \varphi_a 
 + \frac{\lambda}{4!N} \, (\varphi_a \varphi_a)^2 \, .
\eeq
One can classify the various contributions to the 2PI effective action 
according to their scaling with $N$. More generally, one can write
the sum of 2PI diagrams as
\beqn
\Gamma_2[\phi,G]= \Gamma_2^{\rm LO}[\phi,G]
          + \Gamma_2^{\rm NLO}[\phi,G]
          + \Gamma_2^{\rm NNLO}[\phi,G]
          + \ldots \nonumber
\eeqn
where $\Gamma_2^{\rm LO}\sim N$, $\Gamma_2^{\rm NLO}\sim 1$, 
$\Gamma_2^{\rm NNLO}\sim 1/N$ and so on, such that each following 
contribution is suppressed by an additional power of $1/N$. 
The leading order contribution corresponds to only one diagram 
and reads\cite{Cornwall:1974vz}
\beq
{\Gamma_2^{\rm LO}[G] = - \frac{\lambda}{4! N} 
  \int_x G_{aa}(x,x) G_{bb}(x,x) }\, .  
\label{LOcont}
\eeq
Here we use the notation 
$\int_x \equiv \int_\C {\rm d}x^0 \int {\rm d}{\bf x}$. The full 
next-to-leading contribution contains an infinite series of 2PI diagrams
which can be summed:\cite{Berges:2000ur,Aarts:2002dj}
\begin{figure}[b]
 \centerline{\epsfxsize=8.3cm\epsffile{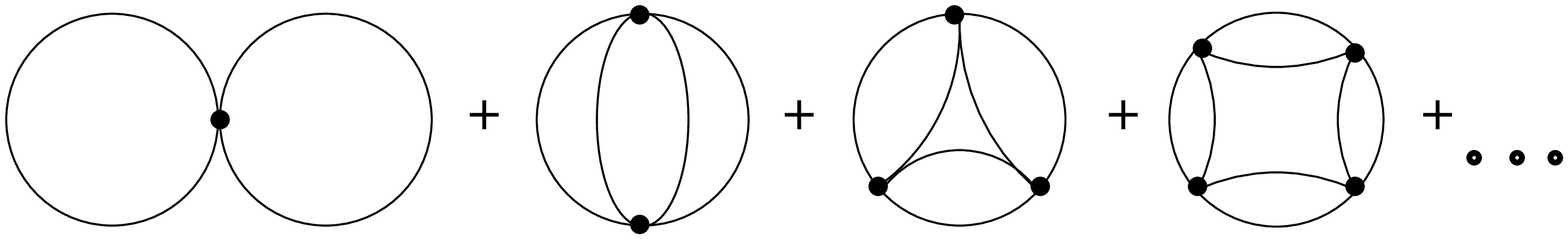}}
 \vspace{.3cm}
 \centerline{\epsfxsize=6.7cm\epsffile{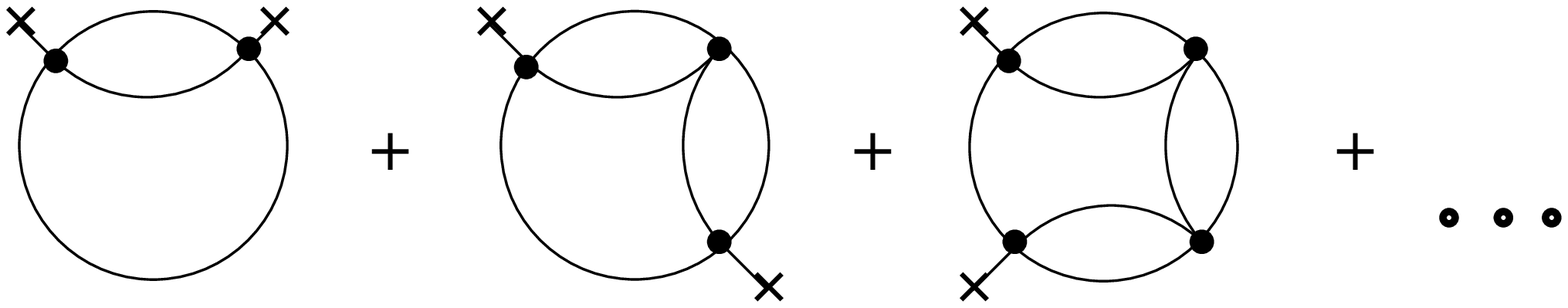}}
 \caption{\label{fig:SSBNLO} The complete NLO contribution to $\Gamma_2$. 
 The dots indicate that each diagram is obtained from the previous 
 one by adding another ``rung'' with two propagator lines at each vertex.
 The crosses denote field insertions.}
\end{figure}
\beq
 \Gamma_2^{\rm NLO}[\phi,G] = \frac{i}{2} \Tr_\C  \mbox{ln} [\, {\bf B}(G)\, ] 
 + \frac{i\lambda}{6N} \int_{xy} {\bf I}(x,y;G) \phi_a(x) G_{ab}(x,y) \phi_b (y) 
 \, ,
\label{NLOcont}
\eeq
where 
\beqn
\label{Feq}
 {\bf B}(x,y;G) = \delta_{\C}(x-y)
 + i \frac{\lambda}{6 N} G_{ab}(x,y)G_{ab}(x,y) \, ,
\eeqn
and the function ${\bf I}(x,y;G)$ sums an infinite series of ``rungs''
(cf.~Fig.~\ref{fig:SSBNLO}):  
\beqn
\label{ieqab}
 {\bf I} (x,y;G) = \frac{\lambda}{6 N} G_{ab}(x,y) G_{ab}(x,y)
 - i \frac{\lambda}{6 N} \int_{z} {\bf I}(x,z;G)
 G_{ab}(z,y) G_{ab}(z,y) \, .
\eeqn
The 2PI $1/N$-expansion provides a controlled nonperturbative approximation
scheme. It can be in particular applied to 
critical phenomena near continuous phase transitions, or to
nonperturbatively large densities as described 
below\cite{Berges:2002cz,Berges:2002wt}.
Instead, for sufficiently dilute systems and weak couplings 
a 2PI coupling-expansion can be appropriate. The lowest non-trivial 
order then contains all diagrams with topology as given by the first 
graph in the upper and lower series presented in Fig.~\ref{fig:SSBNLO}.

\begin{figure}[h]
\vspace*{-0.5cm}
 \centering
 \begin{minipage}[c]{.45\textwidth}
  \centering
  \centerline{\epsfxsize=1.5cm\epsffile{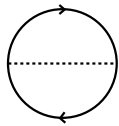}}
 \end{minipage}
 \begin{minipage}[c]{.45\textwidth}
  \centering
  \caption{\label{fig:2loop} Two loop contribution to $\Gamma_2$. 
  The solid and 
  dashed lines represent the full fermionic ($D$) and bosonic 
  ($G$) propagators, respectively.}
 \end{minipage}
\end{figure}

The systematic expansion schemes discussed above can be 
straightforwardly applied to fermionic theories as 
well\cite{Cornwall:1974vz,Berges:2002wr}. As an example, 
consider a theory with two fermion flavors coupled to 
an $O(4)$--vector of scalar fields $\varphi_a \equiv (\sigma,\vec\pi)$
via a chirally invariant interaction with Yukawa coupling $g$:
\beq
\label{potential:Fermion}
 V(\varphi;\bar\psi,\psi)  = \frac{1}{2} m^2_0 \,\,  (\sigma^2 + \pi^2) 
 - g \, \bar{\psi} \left[\,\sigma 
 + i\gamma_5 \,\vec\tau \cdot \vec\pi \,\right] \psi \, .
\eeq
This is a simplified version of the linear sigma model, with no scalar
self-interaction, which is sufficient to obtain 
thermalization as is described below\cite{Berges:2002wr}. 
The 2PI effective action at lowest nontrivial 
order in a systematic coupling expansion then contains only the two-loop
graph shown in Fig.~\ref{fig:2loop}. 

\vspace*{-0.2cm}
\noindent
For discussing nonequilibrium results it is useful to
rewrite the propagators in terms of the spectral function $\rho$ 
and the statistical function $F$\cite{Aarts:2001qa,Berges:2001fi}:
\beq
G (x,y) = F (x,y) -\frac{i}{2} \rho (x,y)\, {\rm sign}_\C(x^0-y^0)
\eeq
and equivalently for the fermion propagator $D$.
We emphasize that out of equilibrium $F$ and $\rho$ are 
{\em independent} functions. However, in thermal equilibrium both functions 
are related by the fluctuation-dissipation
relation, which can be written in Fourier space\cite{Aarts:2001qa,Berges:2001fi}:
\beq
F^{\rm (eq)} (\omega,\vec{p}) = -i
\Big(n_{\rm BE}(\omega)+\frac{1}{2} \Big) \, 
\rho^{\rm (eq)} (\omega,\vec{p}) \, ,
\label{BE}
\eeq
where $n_{\rm BE}(\omega)=(e^{\omega/T}-1)^{-1}$ denotes the Bose-Einstein
distribution with temperature $T$. Starting far from equilibrium, 
Eq.~(\ref{BE}) can be used
to test thermalization at sufficiently late times\cite{Berges:2002wr} 
as is discussed below.

\vspace*{-0.1cm}
\section{\mbox{Comparison of LO, NLO and exact results}}

{\bf\em LO vs.\ NLO:}
Why is it crucial to go beyond leading-order large-$N$ or
Hartree approximations? We consider
for a moment the LO contribution, Eq.~(\ref{LOcont}), 
or the Hartree approximation which also includes the first graph of Fig.\ 
\ref{fig:SSBNLO}. This takes into account space-time dependent mass 
corrections and neglects, in particular, direct scattering. An important 
consequence is the appearance
of an infinite number of conserved quantities, which are not present
in the fully interacting or finite-$N$ theory. These can be written
as\cite{Berges:2001fi,Boyanovsky:vi} 
\beq 
 \Big[F(t,t';p)\, \partial_{t}\partial_{t'} F(t,t';p)
 - \left(\partial_{t} F(t,t';p)\right)^2\Big]^{1/2}_{t=t'}
 \equiv  n_0(p)+\frac{1}{2}\, , 
\label{LOpartnr}
\eeq
where we have considered a spatially homogeneous situation
with $F(t,t';p)$ denoting the spatial Fourier transform.
These additional constants of motion can have a substantial impact
on the time evolution, since they strongly constrain the allowed
dynamics. 
\begin{figure}[b]
 \centerline{\epsfxsize=5.2cm\epsffile{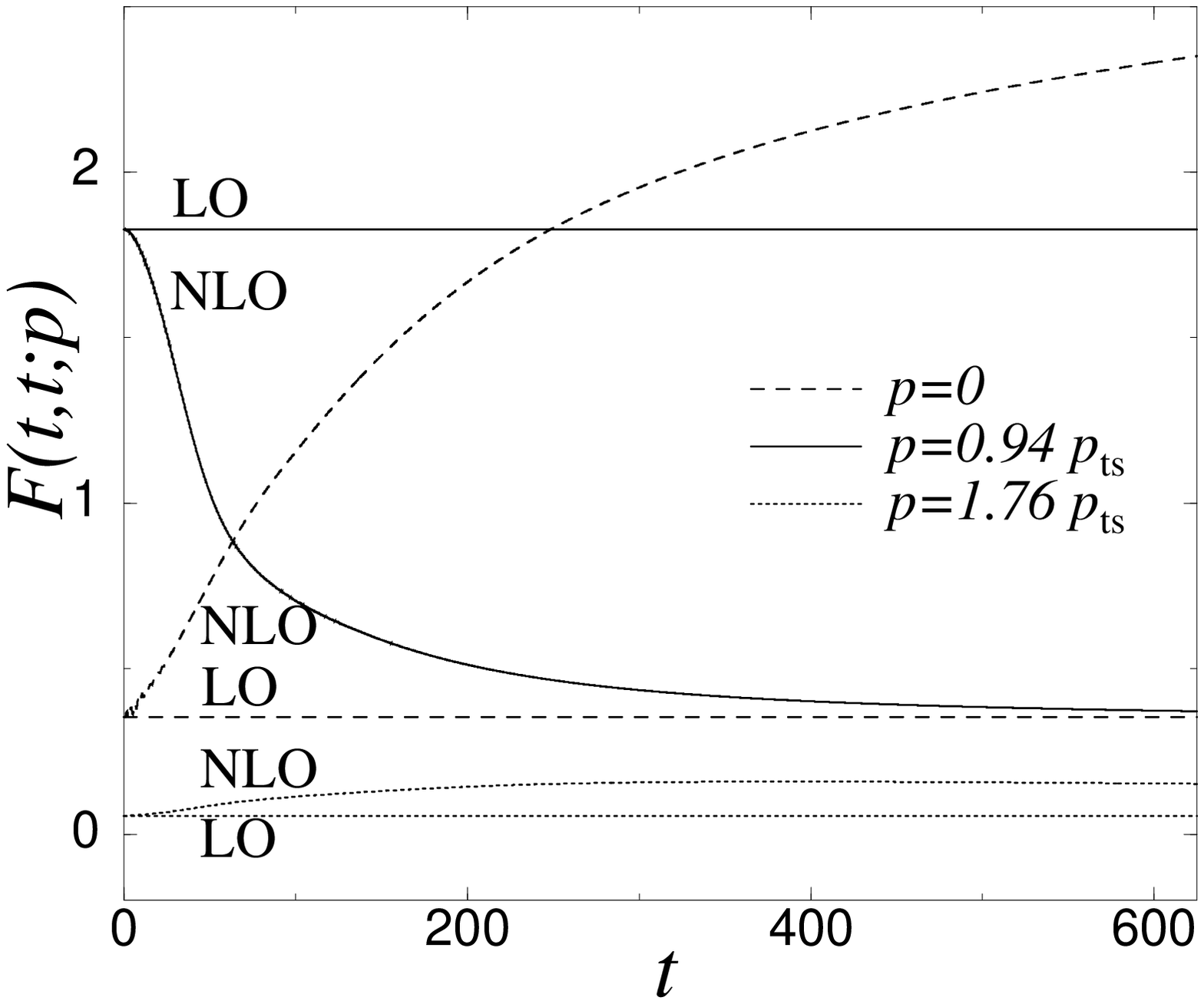}
 \hspace{.5cm}
 \epsfxsize=5.4cm\epsffile{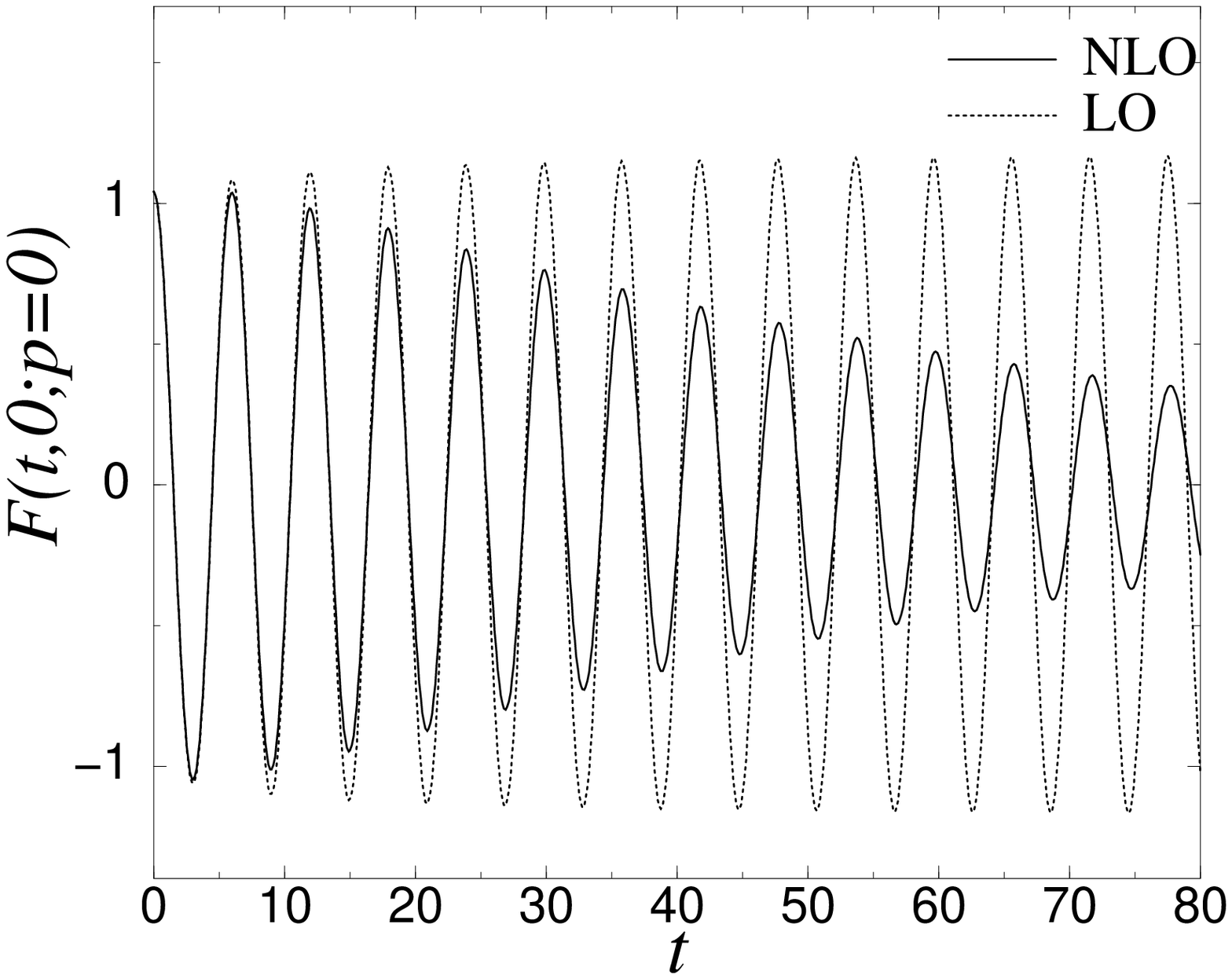}}
\vspace*{-0.2cm}
 \caption{\label{fig:NLOFM} 
Comparison of results at LO and at NLO in the 2PI $1/N$-expansion ($N=10$).
{\bf Left:} {\em Equal-time} two-point function
$F(t,t;p)$ for different momenta as a function of time
for ``tsunami'' initial conditions.
{\bf Right:} {\em Unequal-time} two-point function $F(t,0;p=0)$ as a function
of time following a sudden ``quench''.  
(All in units of the initial-time mass.)} 
\end{figure}
As a characteristic example, 
one may consider the time evolution for 
so-called ``tsunami''\cite{Tsunami} initial conditions with
a vanishing field expectation value: A 
large initial particle number density in a narrow range around a 
characteristic momentum $p = \pm p_{\rm ts}$. Here we consider first
$1+1$ dimensions, where the initial condition 
is reminiscent of two colliding ``wave packets'' 
with opposite and equal momentum. The left graph of 
Fig.~\ref{fig:NLOFM} shows the {\em equal-time} two-point function
$F(t,t;p)$ as a function of time both for the LO and NLO 
approximation\cite{Berges:2001fi}. 
One observes that at NLO the highly populated modes around 
$p_{\rm ts}$ get quickly depopulated, while the 
density for low momentum modes increases. 
In contrast, at LO the evolution of the equal-time correlator is 
trivial. This reflects the fact that at LO or for Hartree approximations,
there exists a well-defined particle number for each momentum mode,
which does not change in time:\cite{Berges:2001fi} 
$n_0(p)$ of Eq.~(\ref{LOpartnr}). 
This is in sharp contrast to the NLO approximation or the fully 
interacting real scalar theory, where there is no conserved particle number
and the LHS of (\ref{LOpartnr}) depends on time.
Similarly strong qualitative differences can also be observed
for {\em unequal-time} correlation functions.
As an example, we show on the right of Fig.~\ref{fig:NLOFM} the
two-point correlation with the initial time, $F(t,0;p=0)$, following
a sudden ``quench''\cite{Berges:2001fi}.
After some oscillations one finds that at NLO the unequal-time
correlator quickly approaches an exponentially damped behavior.  
In contrast, at LO no effective suppression is observed. This
reflects the fact that at LO the presence of additional constants
of motion keep the details about the initial conditions.\\[-0.3cm] 

\begin{figure}[b]
 \centerline{\epsfxsize=5.3cm\epsffile{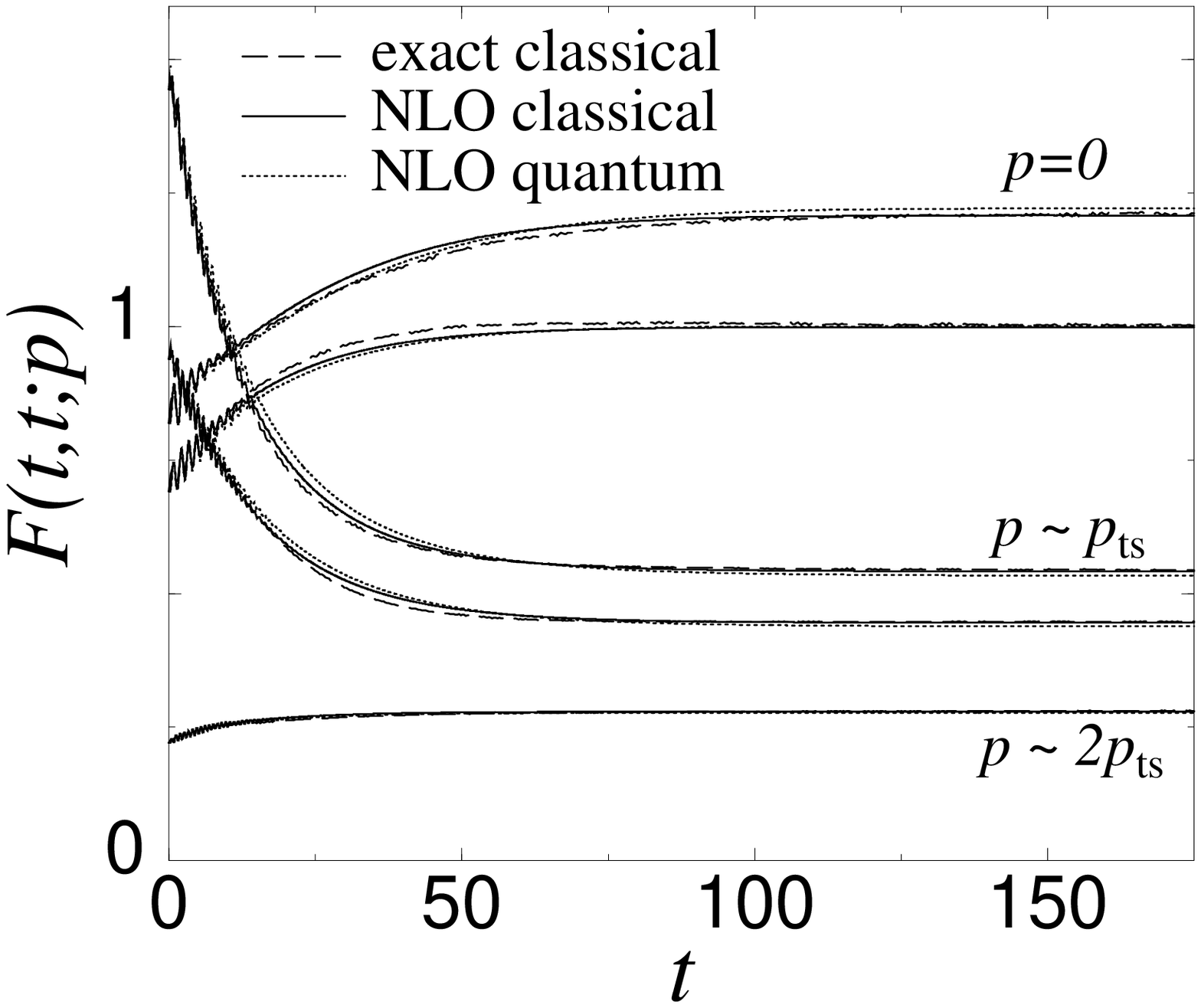}
 \hspace{.5cm}
 \epsfxsize=5.4cm\epsffile{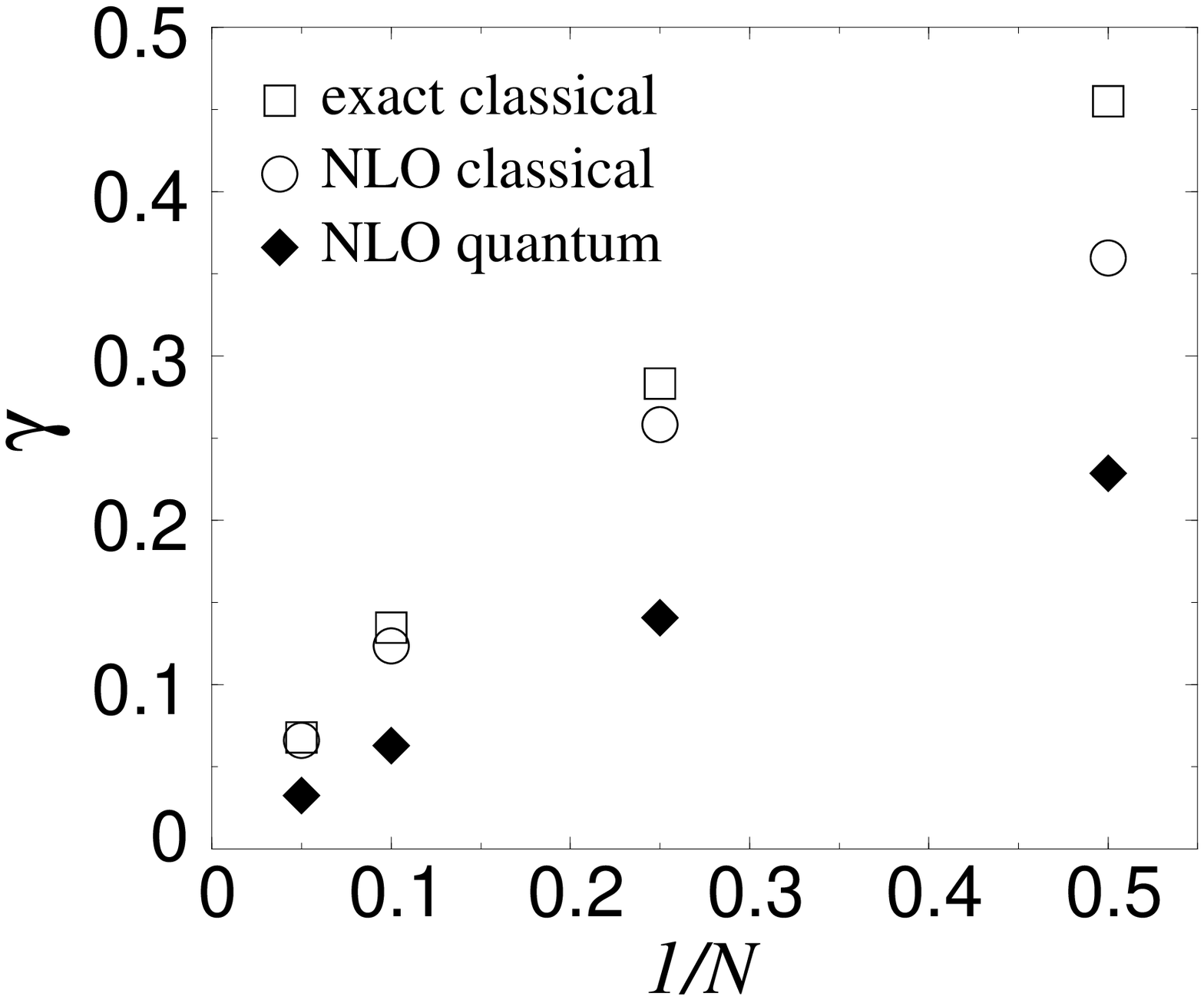}}
\vspace*{-0.2cm}
 \caption{\label{fig:classfig} Comparison between exact (MC) results 
 and results from the 2PI $1/N$--expansion at NLO both for the classical 
 and the quantum field theory. {\bf Left:} Time
 evolution of the equal-time correlation function for an initial condition 
 similar as for Fig.~\ref{fig:NLOFM} with high occupation numbers ($N=4$). 
 {\bf Right:} Damping rate, extracted 
 from the unequal-time correlation function (cf.~Fig.~\ref{fig:NLOFM}, right). 
 Here the initial conditions are characterized by low occupation 
 numbers so that quantum effects become sizeable.}
\end{figure}
{\bf\em NLO vs.\ exact results:}
In view of the substantial changes encountered by going from LO
to NLO for finite $N$, one wonders what happens at NNLO or beyond. 
One can rigorously answer this question in a well-defined limit: 
Nonequilibrium dynamics in {\em classical$\,$} statistical field 
theory can be solved {\em exactly$\,$} up to controlled statistical 
errors,  using numerical integration and Monte Carlo 
techniques\cite{Aarts:2001wi,LOinh,Borsanyi:2000pm,Aarts:2001yn,Blagoev:2001ze}.
The 2PI $1/N$-expansion can be equally well implemented in
the classical as in the quantum case\cite{Aarts:2001yn}. Therefore, in the
classical field limit one can compare with results including 
{\em all} orders in $1/N$. In particular, for
increasing occupation numbers per mode the classical and
the quantum evolution can be shown to approach each other,
if the same initial conditions are applied\cite{Aarts:2001yn}:
For sufficiently high particle number densities one
can strictly verify how rapidly the $1/N$ series
converges. 

Fig.~\ref{fig:NLOFM} shows a comparison
of exact and NLO results\cite{Aarts:2001yn}. 
One observes that the NLO result follows closely the
exact solution already for moderate values of $N$.
This concerns both the far-from-equilibrium dynamics at early times
as well as the late-time behavior close to equilibrium.
The 2PI $1/N$-expansion exhibits very good accuracy 
already in lowest nontrivial order!
We emphasize that also for low densities, where the classical field
approximation provides not a good description for the
quantum theory (cf.~Fig.~\ref{fig:NLOFM}), 
a coupling-expansion to three-loop and a $1/N$-expansion to NLO 
give very similar results for 
sufficiently weak coupling and not too small $N \gtrsim 2$.
One is lead to conclude that, once the spurious conserved quantities 
of LO or Hartree approximations are removed by interactions, 
the results are rather independent of the details of the approximation.\\[-0.3cm]

{\bf\em $1/N$-expansion in the limit $N=1$:} 
The agreement between NLO and exact results is 
found to be rather good for values as small as $N \gtrsim 2$   
(cf.~Fig.~\ref{fig:classfig})\cite{Aarts:2001yn}. 
However, in the limit $N=1$ subleading contributions 
in $1/N$ are no longer suppressed and, in particular, 
the $1/N$-expansion at NLO  
exhibits a reduced damping rate\footnote{We note that
this can be understood from the fact that the three-loop 
contribution to the 2PI effective action (which gives the lowest order 
contribution to the damping rate in a coupling expansion) is 
$\sim(\lambda/N)^2[1+2/N]$. Comparing this with the  
corresponding three-loop contribution at NLO in the 
$1/N$-expansion\cite{Berges:2001fi,Aarts:2002dj} for $N=1$, 
one finds a factor of three less for the latter 
(cf.~Sect.~\ref{sect:app}).}. The case $N=1$ 
has been investigated in classical field theory
in Ref.~\refcite{Blagoev:2001ze} 
for initial conditions with zero and nonzero field $\phi$.
An approximation, which includes part of the NNLO corrections, 
such as BVA\cite{Mihaila:2000sr}, is found to give a better
damping rate for the field. However, a worse result is obtained
for the two-point function, which reflects the absence of an 
expansion parameter for $N=1$.
For weak interactions a systematic expansion
can be based on the 2PI coupling-expansion~for~\mbox{$N=1$}.\\[-0.3cm] 

\begin{figure}[t]
\vspace*{-0.3cm}   
 \centerline{\epsfxsize=6.cm\epsfbox{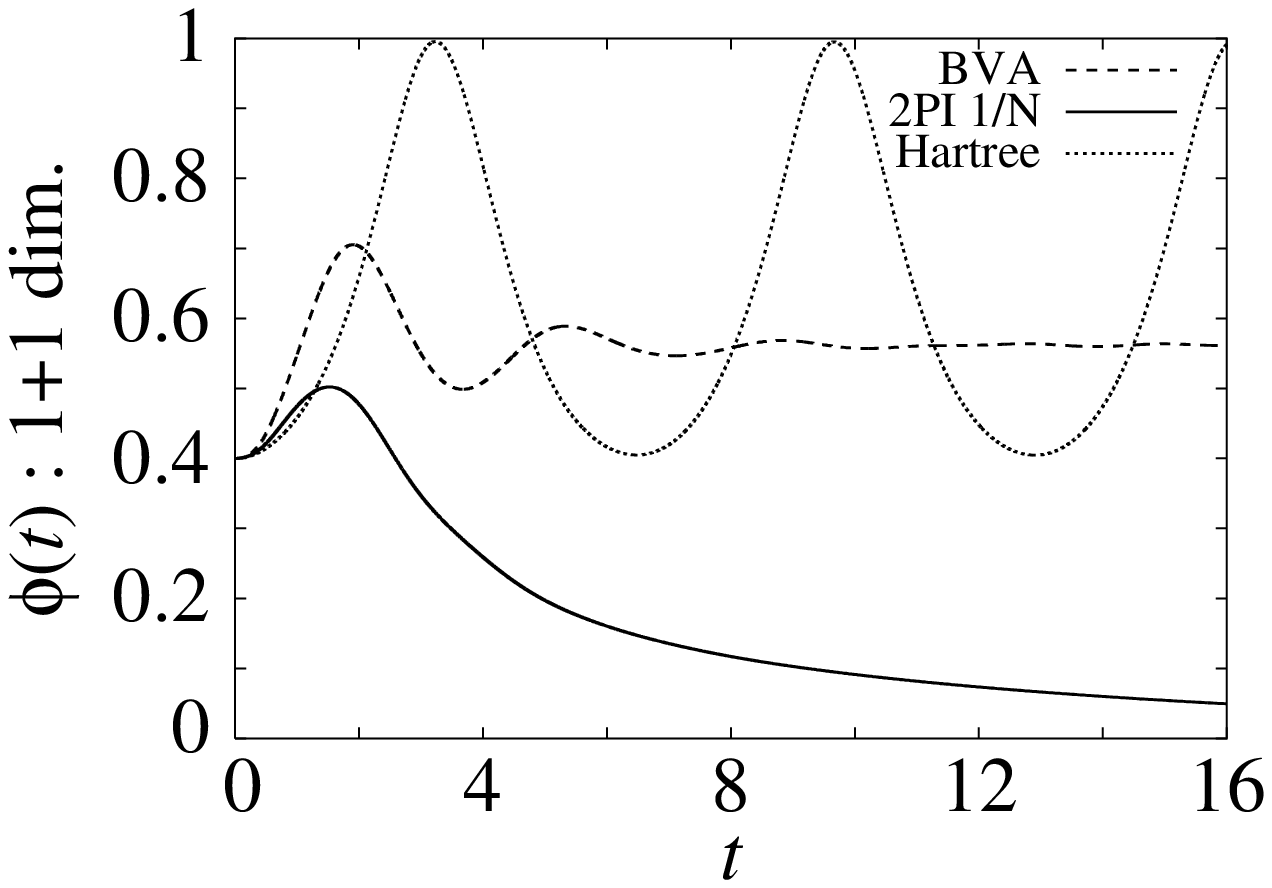}
 \epsfxsize=5.5cm\epsfbox{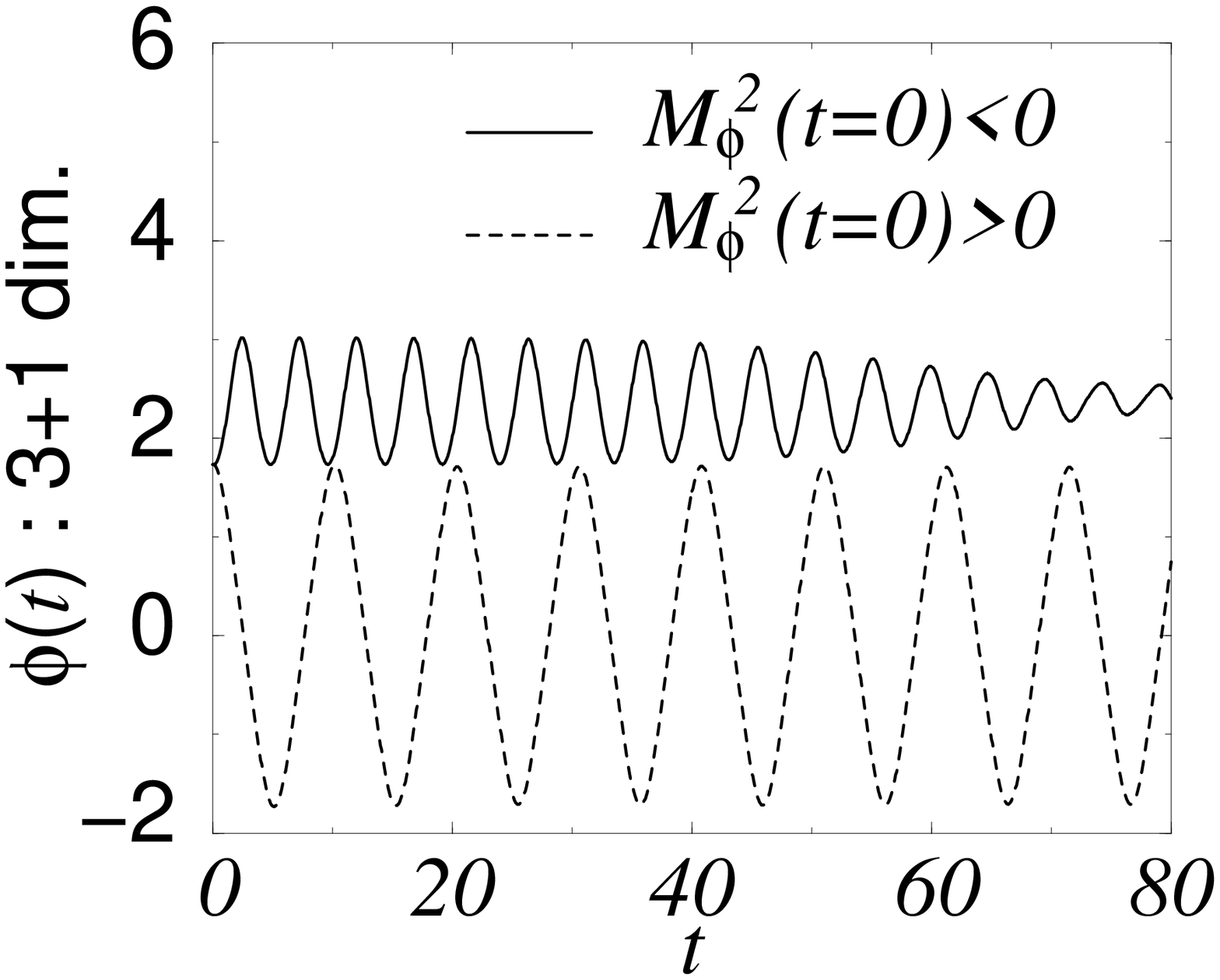}}
\vspace*{-0.5cm}   
 \caption{\label{fig:ssb} Examples of nonequilibrium time evolutions
of the field expectation value $\phi(t)$ for one (left$^{10}$) and for three
(right) spatial dimensions using different approximations. 
The 2PI $1/N$-expansion (here for the limit $N=1$) 
correctly describes the absence of a spontaneously broken phase 
in one spatial dimension, with $\phi(t) \to 0$ at late times. 
In contrast, spontaneous symmetry breaking can be observed
at NLO in $3+1$ dimensions.}
\end{figure}

{\bf\em Spontaneous symmetry breaking:}
It is a well-known fact that there is no spontaneous 
symmetry breaking in one spatial dimension at nonzero 
temperature or energy density. In contrast, 
LO or Hartree approximations can exhibit a 
phase transition independent of the number of dimensions. 
For nonequilibrium, 
Cooper at al.\ have pointed out in Ref.~\refcite{Cooper:2002qd}
that the 2PI $1/N$--expansion at NLO 
shows no spontaneous symmetry breaking in one spatial dimension,
thus curing the problem of mean-field type approximations.
As an example, the left graph of Fig.~\ref{fig:ssb} (taken 
from Ref.~\refcite{Cooper:2002qd}) shows their results
for the time evolution of the field expectation value $\phi(t)$. 
One observes that, while the Hartree approximation oscillates around 
a nonzero field value, the NLO result tends to the symmetric phase 
with $\phi = 0$. We emphasize, however, that the latter results have been
obtained from a $1/N$-expansion in the uncontrolled limit $N=1$.
This becomes obvious once e.g.\ a part but not all of the NNLO 
contributions are taken into account, as denoted by BVA in 
Fig.~\ref{fig:ssb}, where again one finds spontaneous
symmetry breaking in one spatial dimension. More recent results, 
demonstrating the absence of a nonequilibrium phase transition in 
one spatial dimension have been obtained from a loop-expansion in a 
related (2PPI) approximation scheme\cite{Baacke:2002ee}. 
It is interesting
that the systematic $1/N$-expansion is found to give qualitatively 
correct results even for $N=1$. Here we show for the first time that
the 2PI $1/N$-expansion exhibits spontaneous symmetry breaking in 
$3+1$ dimensions. The right graph of Fig.~\ref{fig:ssb} 
shows the evolution for a sufficiently negative and a positive initial mass
parameter $M_{\phi}^2$, which leads to a nonzero and a vanishing
field expectation value respectively.  

\vspace*{-0.1cm}
\section{Applications}
 
\subsection{\label{largefluc} Parametric resonance in quantum field theory}

In classical mechanics 
parametric resonance is the phenomenon of resonant amplification of 
the amplitude of an oscillator having a time-dependent periodic frequency.
In the context of quantum field theory a similar phenomenon describes 
the amplification of quantum fluctuations, which can be interpreted as 
particle production. It provides an important building block for our 
understanding of the (pre)heating of the early universe at the end of an 
inflationary period\cite{Kofman:1994rk}, and may also be operative 
in various situations 
in the context of relativistic heavy-ion collisions\cite{Dumitru:2000in}. 

The phenomenon of parametric resonance provides a paradigm for the dynamics of
quantum fields at nonperturbatively large densities\footnote{Another 
important example, where nonperturbatively high densities occur,
is provided by the physics of the color glass condensate\cite{CGC} in QCD.}.
It is a far-from-equilibrium phenomenon involving the production of particles 
with densities inversely proportional to the coupling. For this reason, it 
has been much studied in the classical field approximation, which has
long been the only available quantitative approach\cite{Khlebnikov:1996mc}. 
Studies in quantum field theory have so far been limited to linear 
or mean-field type approximations\cite{Kofman:1994rk}. 
\begin{figure}[t]
 \centerline{\hspace*{-1.5cm}\epsfxsize=6.2cm\epsfbox{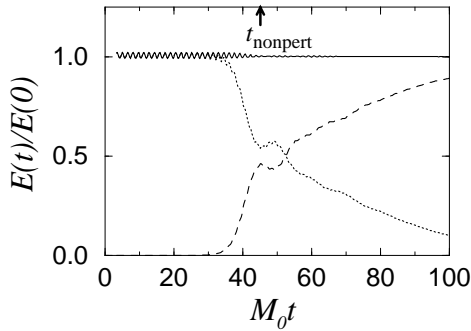}}   
\vspace*{-0.2cm}
 \caption{\label{fig:energy} Total energy (solid line) and 
classical field energy (dotted line) as a function of time for
$\lambda =10^{-6}$. The dashed line represents the fluctuation part,
showing a transition between a classical field and a fluctuation dominated 
regime.}
\end{figure}

Using the methods described in the previous sections, 
recently the first quantum field 
theoretical study of this phenomenon
beyond LO or Hartree approximations
has been performed\cite{Berges:2002cz}. To deal with the 
nonperturbatively large densities, one can employ the 2PI $1/N$--expansion
at NLO, in the presence of a nonvanishing field (the ``inflaton'')
in $3+1$ dimensions.  

\begin{figure}[b]
 \centerline{\hspace*{-1.5cm}\epsfxsize=6.9cm\epsfbox{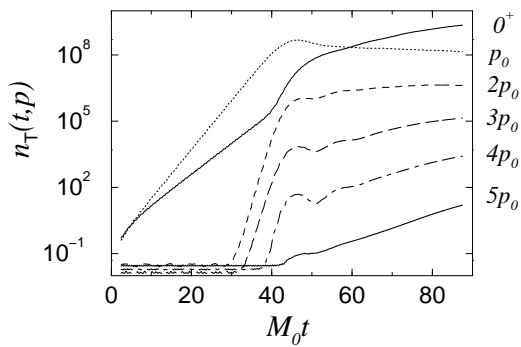}}   
 \caption{\label{fig:number} Effective particle number density for the 
 transverse modes as a function of time for various momenta 
 $0 \le p \le 5 p_0$ 
 and $\lambda =10^{-6}$. At early times, modes with momentum 
 $p \simeq p_0$ are  
 exponentially amplified with a rate $\sim 2 \gamma_0$ via the mechanism of 
 parametric resonance. Due to nonlinearities, one observes 
 subsequently an enhanced growth
 with rate~$\sim 6\gamma_0$ for a broad momentum range.}
\end{figure}

We consider\cite{Berges:2002cz} a system that is initially 
in a pure quantum state, characterized by
a large field amplitude $\phi_a(t=0)\sim 1/\sqrt \lambda$ and 
comparatively small quantum fluctuations, described here by employing 
vanishing particle numbers at initial time. Fig.\ \ref{fig:energy} 
provides an overview of the dynamics: At early times, the total energy 
of the system is dominated by the contribution from the large field 
amplitude, which subsequently decays into particles, giving rise to
a fluctuation dominated regime at late times. More precisely, the early 
time coherent oscillations of the field trigger an exponential amplification 
of quantum fluctuations, corresponding to explosive particle production in a 
narrow range of momenta centered at $p=p_0$: this is parametric 
resonance.\cite{Kofman:1994rk} Later on, nonlinear interactions
between field modes cause this amplification to propagate to higher momenta.
More specifically, the initially amplified modes act as a source for other
modes. This source-induced amplification results in an enhanced particle
production in a broad momentum range. This is illustrated in Fig.\ 
\ref{fig:number}, where the effective particle number is shown
for various momenta as a function of time from a numerical
solution.  

We emphasize that, using the above 2PI techniques, one can obtain 
the main features of the nonlinear dynamics up to $t_{\rm nonpert}$ 
also from an analytical solution of the evolution 
equations\cite{Berges:2002cz}. In particular, in order for 
the NLO effects to be small compared
to LO dynamics at early times, one needs values of $N$ as large as
$N \gtrsim 1/\lambda$. The late-time dynamics for 
$t \gg t_{\rm nonpert}$ requires
the numerical solution of the NLO equations. For the  
employed small couplings, this involves extremely large times which has
not been undertaken up to now.   
It has recently been shown using classical field theory
methods that the very slow subsequent evolution is characterized by turbulent 
behavior\cite{Micha:2002ey}. However, to describe the late-time approach 
to quantum thermal equilibrium with a Bose-Einstein distributed
particle number, one needs to go beyond the classical field 
approximation.

\subsection{\label{sec:fermion} Nonequilibrium chiral quark-meson model}

We consider a theory involving two Dirac fermion
flavors (``quarks'') coupled to a scalar
$\sigma$--field and a triplet of pseudoscalar ``pions''
as described by Eqs.~(\ref{potential:largeN}).
Such type of linear $\sigma$-models, which implement the approximate
chiral $SU_L(2)\times SU_R(2)$ symmetry of QCD,
can be a useful first guide for the behavior of 
strongly interacting matter relevant for the 
dynamics of relativistic heavy-ion collisions.
\begin{figure}[b]
\vspace*{-0.6cm}
 \centerline{\epsfxsize=7.cm\epsfbox{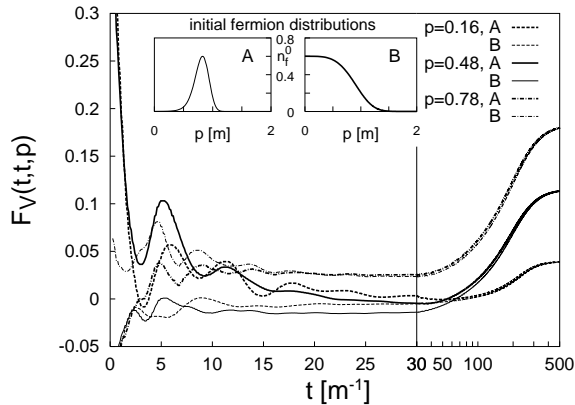}}   
\vspace*{-0.2cm}
 \caption{\label{fig:join} Time evolution of fermionic 
 equal-time two-point functions for various momenta and for two 
 different initial conditions $A$ and $B$. The corresponding initial particle
 number distributions, shown in the insets, are chosen such that the total
 energy is the same for both cases. (In units of the scalar 
{\em thermal} mass $m$.)}
\end{figure}

Recently\cite{Berges:2002wr} the far-from-equilibrium time evolution and 
subsequent thermalization has been described for
this model at lowest nontrivial 
order in the 2PI coupling-expansion, 
corresponding to the two-loop 
graph presented in Fig.\ \ref{fig:2loop}. 
Starting from various different far-from-equilibrium 
initial conditions, one finds a universal late-time behavior
that is only determined by the expectation value of the
initial energy density. In particular, one is able to
observe the approach to Bose-Einstein and Fermi-Dirac 
distributions at sufficiently late times. Similar 
studies of the late-time behavior have previously only 
been performed for purely scalar theories in $1+1$ 
dimensions\cite{Berges:2000ur,Berges:2001fi}. 
  
\begin{figure}[t]
\vspace*{-0.2cm}
 \centerline{
 \epsfxsize=7.cm\epsfbox{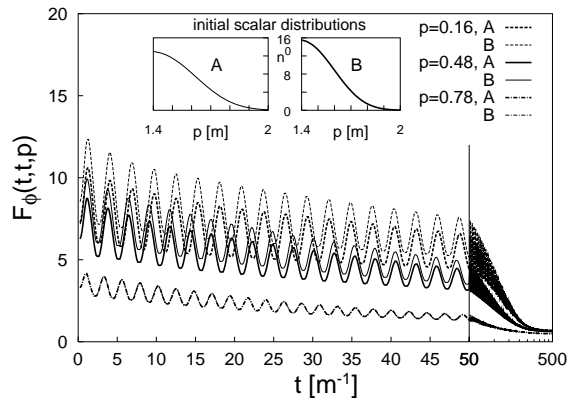}}   
\vspace*{-0.2cm}
 \caption{\label{fig:join2} Same as in Fig.~\ref{fig:join} but for the scalars.}
\end{figure}

As an example, in Figs.~\ref{fig:join} 
and \ref{fig:join2} the statistical two-point functions
for the fermions, $F_V(t,t;p)$, and for the scalars,
$F_{\phi}(t,t;p)$, are shown for two very different nonequilibrium
initial conditions (cf.~Ref.~\refcite{Berges:2002wr} for details). 
One observes that the time evolution becomes soon rather   
insensitive to the details of the initial condition. 
The time for the effective loss of initial conditions is
well described by the inverse damping rate obtained from
the respective unequal-time two-point function (cf.~the example in
Fig.~\ref{fig:NLOFM}). In contrast, 
this time scale does not characterize the late-time
behavior. For the latter, one finds, to very good approximation, an 
exponential relaxation of each mode to their universal 
late-time values. For the employed initial conditions
the thermalization rate is somewhat larger than the damping rate.
Though quantitatively very different, a similar qualitative behavior
has been previously observed for scalar theories in $1+1$ 
dimensions\cite{Berges:2000ur,Berges:2001fi}.

\begin{figure}[t]
 \centerline{
 \epsfxsize=7.cm\epsfbox{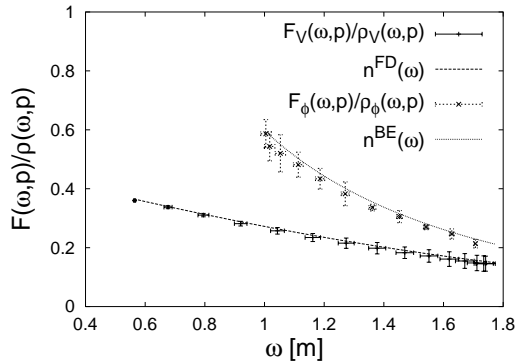}}   
 \caption{\label{fig:join3}
The late-time ratio of the statistical two-point function  
and the spectral function in frequency space, both for
fermions ($F_V/\rho_V$) and for scalars ($F_\phi/\rho_\phi$).  
For thermal equilibrium the quotient has to correspond to the Bose-Einstein
(BE) distribution for scalars and to the Fermi-Dirac (FD) 
distribution for fermions. 
The BE/FD distributions are displayed by the continuous curves 
with the same temperature $T=0.94m$.}
\end{figure}

At sufficiently late times one can explicitly demonstrate 
the approach to quantum thermal equilibrium\footnote{We emphasize 
that thermal equilibrium cannot 
be reached on a fundamental level from time-reversal invariant 
evolution equations at any finite time. The results 
demonstrate\cite{Berges:2000ur,Berges:2001fi,Berges:2002wr} that
thermal equilibrium can be approached very closely at sufficiently 
late time, without again deviating from it for practically accessible
times.}: 
Out of equilibrium, the spectral and statistical two-point functions are 
completely independent in general. However, if thermal equilibrium
is approached, they have to become related by the fluctuation-dissipation
relation at late times as stated in Eq.~(\ref{BE}).
For sufficiently late times one observes that
the correlators become approximately homogeneous in time and a Fourier
transform with respect to $t-t'$ can be performed to very good 
approximation. The result is shown in Fig.~\ref{fig:join3}.

This work provides
an important first step for a quantitative description of
realistic theories with fermions in $3+1$ dimensions,
with important phenomenological applications\cite{Serreau:CERN} to   
the out-of-equilibrium chiral phase 
transition, or the physics of baryogenesis in 
the early universe\cite{Prokopec}.

\section{Outlook}

The 2PI generating functional for Green's functions provides 
a powerful technique for nonequilibrium physics. Systematic 
approximation schemes can describe 
far-from-equilibrium dynamics as well as subsequent thermalization 
from ``first principles''. Substantial progress has been achieved 
along these lines in recent years and, by now, the methods 
provide a practical means for quantitative descriptions 
of realistic scalar and fermionic quantum field theories. 
Apart from the discussed applications motivated by
high--energy particle physics and cosmology, we emphasize that the
same techniques can be applied to condensed matter systems. As an 
important example, the dynamics of Bose--Einstein condensation can be 
addressed along very similar lines. 

One important issue, which we have not discussed here, concerns the 
renormalizability of approximations based
on truncations of the 2PI effective action. To prove this is a 
nontrivial task because of the fully self-consistent character of these 
approximations. Recent work has shown that, in the context of scalar 
field theory, approximations based on a systematic loop-expansion 
of the 2PI effective action (``$\Phi$-derivable approximations'') 
can be renormalized order by order, 
if the underlying theory is perturbatively 
renormalizable\cite{vanHees:2002js,Reinosa,Braaten:2001vr}.

Another, major open question concerns the description of gauge fields. The 
difficulty is related to the fact that truncations of the 2PI effective 
action in general violate Ward identities. One possibility to avoid this
difficulty is to modify the 2PI-based approximations in such a way as to
enforce Ward identities. Practical examples in this direction have been 
pointed out for nonequilibrium systems\cite{Mottola}.
Another interesting possibility is to try to keep the gauge dependence of
physical results under control. This has recently been investigated 
in Ref.~\refcite{Arrizabalaga:2002hn} and further interesting developments 
are to be awaited.

\vspace*{-0.2cm}
\section*{Acknowledgments}
We thank Gert Aarts, Daria Ahrensmeier, Rudolf Baier, 
Szabolcs Bors\'anyi, J{\"u}rgen Cox, Markus M.~M{\"u}ller
and Christof Wetterich 
for very fruitful collaborations on this topic.

\end{document}